\documentclass[12pt,preprint]{aastex}

\begin{document}

\title{Dark energy and cosmic curvature: Monte-Carlo Markov Chain approach}

\author{Yungui Gong}
 \affil{College of Mathematics and Physics, Chongqing University of Posts and Telecommunications,
Chongqing 400065, China} \email{gongyg@cqupt.edu.cn}
\author{Qiang Wu}
 \affil{GCAP-CASPER, Department of
Physics, Baylor University, Waco, TX 76798, USA}
\email{qiang\_wu@baylor.edu} \and
\author{Anzhong Wang}
 \affil{ Department of Theoretical
Physics, Institute of Physics, the State University of Rio de
Janeiro, Brazil\\
and\\
GCAP-CASPER, Department of Physics, Baylor University,
Waco, TX 76798, USA} \email{anzhong\_wang@baylor.edu}

\begin{abstract}
We use the Monte-Carlo Markov Chain method to explore the dark
energy property and the cosmic curvature by fitting two popular dark
energy parameterizations to the observational data. The new 182 gold
supernova Ia data and the ESSENCE data both give good constraint on
the DE parameters and the cosmic curvature for the dark energy model
$w_0+w_a z/(1+z)$. The cosmic curvature is found to be
$|\Omega_k|\la 0.03$. For the dark energy model $w_0+w_a z/(1+z)^2$,
the ESSENCE data gives better constraint on the cosmic curvature and
we get $|\Omega_k|\leq 0.02$.
\end{abstract}

\keywords{Cosmology: cosmological parameters --- Cosmology:
observations}



\section{Introduction}

The supernova (SN) Ia observations indicate the accelerated
expansion of the Universe \citep{riess98,perlmutter}. The direct and
model independent evidence of the acceleration of the Universe was
shown by using the energy conditions in \citet{gong07b} and
\citet{gong07c}. The driving force of the late time acceleration of
the Universe, dubbed ``dark energy (DE)", imposes a big challenge to
theoretical physics. Although the cosmological constant is the
simplest candidate of DE and consistent with current observations,
other possibilities are also explored due to many orders of
magnitude discrepancy between the theoretical estimation and
astronomical observations for the cosmological constant. For a
review of DE models, see for example,
\citet{review1,review2,review3,review4,review5}.

There are model independent studies on the nature of DE by using the
observational data. In particular, one usually parameterizes DE
density or the equation of state parameter $w(z)$ of DE
\citep{alam04a,alam04b,astier,barger,cardone,mcdp,trc,clarkson,corasaniti,efstathiou,gerke,gong05a,gong05b,gong05c,gong06,
gong07a,gu,huterer,huterer05,ichikawa6,ichikawa,ichikawa7,jbp,jonsson,lee,linder,setare,cooray,flux,flux1,flux2,weller01,weller02,wetterich,zhu}.
Due to the degeneracies among the parameters in the model,
complementary cosmological observations are needed to break the
degeneracies. The Wilkinson Microwave Anisotropic Probe (WMAP)
measurement on the Cosmic Microwave Background (CMB) anisotropy,
together with the SN Ia observations provide complementary data. In
this paper, we use the three-year WMAP (WMAP3) data \citep{wmap3},
the SN Ia data \citep{riess06,essence,essence1} and the Baryon
Acoustic Oscillation (BAO) measurement from the Sloan Digital Sky
Survey \citep{sdss} to study the property of DE and the cosmic
curvature. Two DE models $w(z)=w_0+w_a z/(1+z)$ \citep{mcdp,linder}
and $w(z)=w_0+w_a z/(1+z)^2$ \citep{jbp} are considered. In
\citet{lar}, the authors showed that combining the shift parameters
$R$ and the angular scale $l_a$ of the sound horizon at
recombination appears to be a good approximation of the full WMAP3
data. Wang and Mukherjee gave model independent constraints on $R$
and $l_a$ by using the WMAP3 data, they also provided the covariance
matrix of the parameters $R$, $l_a$ and $\Omega_b h^2$
\citep{wang07}. So we use the shift parameter $R$, the angular scale
$l_a$ of the sound horizon at recombination and their covariance
matrix given in \citet{wang07} instead to avoid using several
inflationary model parameters and calculating the power spectrum.
When the covariance matrix is used, we have six parameters. We use
the Monte-Carlo Markov Chain (MCMC) method to explore the parameter
space. Our MCMC code is based on the publicly available package
COSMOMC \citep{cosmomc}.

The paper is organized as follows. In section II, we give all the
formulae and show the constraint on the cosmic curvature is much
better by using the parameters $R$, $l_a$ and their covariance
matrix than that by using the parameter $R$ only. We also discuss
the effect of the radiation component $\Omega_r$ on $l_a$. In
section III, we give our results. We discuss the analytical
marginalization over $H_0$ in appendix A.

\section{Method}

For the SN Ia data, we calculate
\begin{equation}
\label{chi}
\chi^2=\sum_i\frac{[\mu_{obs}(z_i)-\mu(z_i)]^2}{\sigma^2_i},
\end{equation}
where the extinction-corrected distance modulus
$\mu(z)=5\log_{10}[d_L(z)/{\rm Mpc}]+25$, $\sigma_i$ is the total
uncertainty in the SN Ia data, and the luminosity distance is
\begin{equation}
\label{lumdis}
d_{\rm L}(z)=\frac{1+z}{H_0\sqrt{|\Omega_{k}|}} {\rm
sinn}\left[\sqrt{|\Omega_{k}|}\int_0^z
\frac{dz'}{E(z')}\right],
\end{equation}
here
\begin{eqnarray}
\frac{{\rm sinn}(\sqrt{|\Omega_k|}x)}{\sqrt{|\Omega_k|}}=\left\{\begin{array}{lr}
\sin(\sqrt{|\Omega_k|}x)/\sqrt{|\Omega_k|},& {\rm if}\ \Omega_k<0,\\
x, & {\rm if}\  \Omega_k=0, \\
\sinh(\sqrt{|\Omega_k|}x)/\sqrt{|\Omega_k|}, & {\rm if}\  \Omega_k>0,
\end{array}\right.
\end{eqnarray}
and the dimensionless Hubble parameter is
\begin{equation}
\label{ezdef}
E^2(z)=H^2(z)/H^2_0=\Omega_m(1+z)^3+\Omega_r (1+z)^4+\Omega_k (1+z)^2+\Omega_{DE},
\end{equation}
where $\Omega=8\pi G\rho/(3H^2_0)$, $\rho_r=\sigma_bT_{cmb}^4$,
$\sigma_b$ is the Stefan-Boltzmann constant, the CMB temperature
$T_{cmb}=2.726$K, and $\Omega_{DE}$ is the DE density. Note that the
distance normalization is arbitrary in the SN Ia data, the Hubble
constant $H_0$ determined from the SN data is also an arbitrary
number, not the observed Hubble constant. Therefore we need to
marginalize over this nuisance parameter $H_0$. The parameter $H_0$
is marginalized over with flat prior, the analytical marginalization
method is discussed in Appendix A. For the DE model
\citep{mcdp,linder}
\begin{equation}
\label{lind}
w(z)=w_0+\frac{w_a z}{1+z},
\end{equation}
the dimensionless DE density is
\begin{equation}
\label{deneq}
\Omega_{DE}(z)=(1-\Omega_m-\Omega_k-\Omega_r)(1+z)^{3(1+w_0+w_a)}\exp[-3w_a z/(1+z)].
\end{equation}
For the DE model \citep{jbp}
\begin{equation}
\label{wzeq}
w(z)=w_0+\frac{w_a z}{(1+z)^2},
\end{equation}
the dimensionless DE density is
\begin{equation}
\label{deneq1}
\Omega_{DE}(z)=(1-\Omega_m-\Omega_k-\Omega_r)(1+z)^{3(1+w_0)}\exp\left[3w_a z^2/2(1+z)^2\right].
\end{equation}

For the SDSS data, we add the term
$$\left[\frac{A-0.469(0.95/0.98)^{-0.35}}{0.017}\right]^2$$
to $\chi^2$ \citep{sdss,wmap3}, where the BAO parameter
\begin{equation}
\label{para1}
A=\frac{\sqrt{\Omega_{m}}}{0.35}\left[\frac{0.35}{E(0.35)}\frac{1}{|\Omega_{k}|}{\rm
sinn}^2\left(\sqrt{|\Omega_{k}|}\int_0^{0.35}
\frac{dz}{E(z)}\right)\right]^{1/3}.
\end{equation}

For WMAP3 data, we first add the term
$$\left(\frac{R-1.71}{0.03}\right)^2$$
to $\chi^2$ \citep{wang07}, where the shift parameter
\begin{equation}
\label{shift1}
R=\frac{\sqrt{\Omega_{m}}}{\sqrt{|\Omega_{k}|}}{\rm
sinn}\left(\sqrt{|\Omega_{k}|}\int_0^{z_{ls}}\frac{dz}{E(z)}\right),
\end{equation}
and $z_{ls}=1089\pm 1$.

\begin{figure}
\centering
\includegraphics[width=12cm]{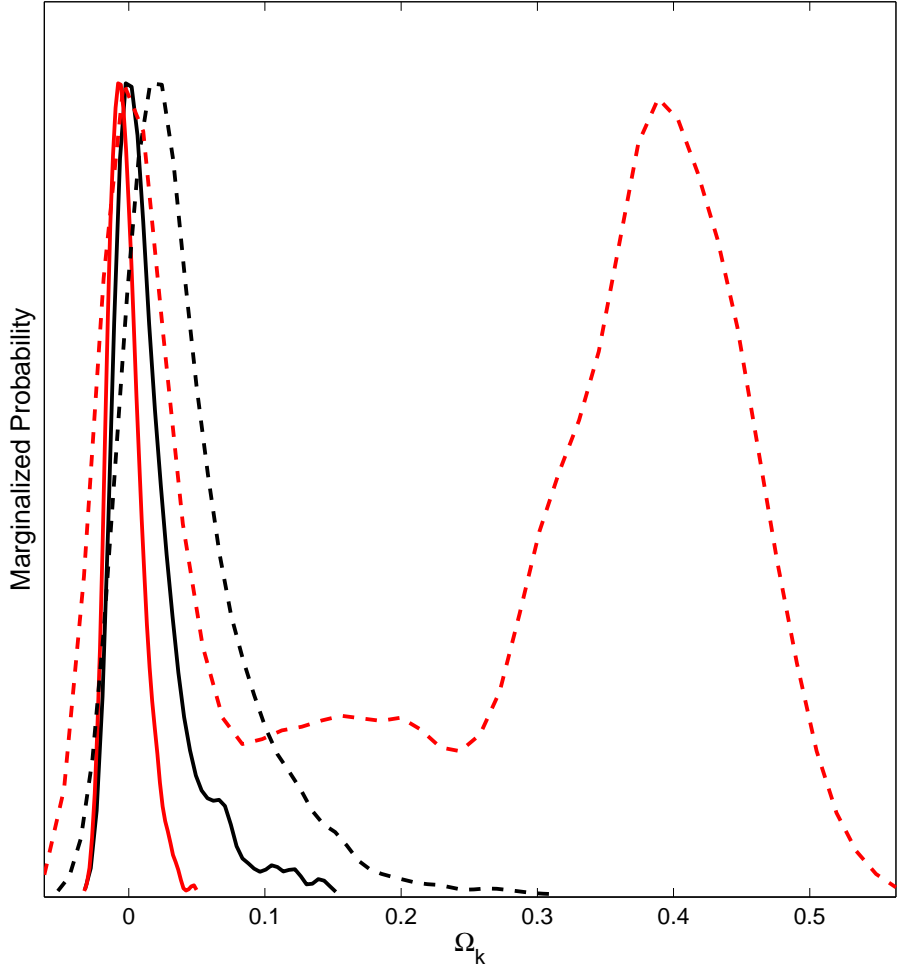}
\caption{The marginalized probabilities of $\Omega_k$. The solid lines denote the results
using the shift parameter $R$, the angular scale $l_a$, and the full covariance matrix. The dashed
lines denote the results using the shift parameter only. The black lines are for
the dark energy model $w_0+w_a z/(1+z)^2$ and the red lines are for the model
$w_0+w_a z/(1+z)$.}
\label{fig1}
\end{figure}

When we fit the DE models (\ref{lind}) and (\ref{wzeq}) to the
observational data, we have four parameters $\Omega_m$, $\Omega_k$,
$w_0$ and $w_a$. The MCMC method is used to explore the parameter
space. The marginalized probability of $\Omega_k$ is shown in Fig.
\ref{fig1}. It is obvious that the cosmic curvature cannot be well
constrained for the DE model (\ref{lind}). As discussed in
\citet{lar} and \citet{wang07}, the combination of the shift
parameter and the angular scale of the sound horizon at
recombination gives much better constraints on cosmological
parameters. So we add the angular scale of the sound horizon at
recombination \citep{wang07}
\begin{equation}
\label{csla}
l_a=\frac{\pi R/\sqrt{\Omega_m}}{\int_{z_{ls}}^\infty dz c_s/E(z)}=302.5\pm 1.2,
\end{equation}
where the sound speed $c_s=1/\sqrt{3(1+\bar{R_b} a)}$,
$\bar{R_b}=315000\Omega_b h^2(T_{cmb}/2.7{\rm K})^{-4}$, $a$ is the
scale factor, and $\Omega_b h^2=0.02173\pm 0.00082$ \citep{wang07}.
To implement the WMAP3 data, we need to add three fitting parameters
$R$, $l_a$ and $\Omega_b h^2$. So we need to add the term $\Delta
x_i {\rm Cov}^{-1}(x_i,x_j)\Delta x_j$ to $\chi^2$, where $x_i=(R,\
l_a,\ \Omega_b h^2)$ denote the three parameters for WMAP3 data,
$\Delta x_i=x_i-x_i^{obs}$ and Cov$(x_i,x_j)$ is the covariance
matrix for the three parameters. Follow Wang and Mukherjee, we use
the covariance matrix for $x_i=(R,\ l_a,\ \Omega_b h^2)$ derived in
\citet{wang07}. Since the covariance matrix for the six quantities
in \citet{wang07} is defined as the pair correlations for those
variables, so each element in the matrix is obtained by
marginalizing over all other variables. Therefore, the covariance
matrix between $x_i$ and $x_j$ is the three by three sub-matrix  of
the full six by six matrix in \cite{wang07}. The marginalized
probability of $\Omega_k$ is shown in Fig. \ref{fig1}. We see that
the cosmic curvature is constrained better with the addition of the
angular scale $l_a$ of the sound horizon at recombination.

\begin{figure}
\centering
\includegraphics[width=12cm]{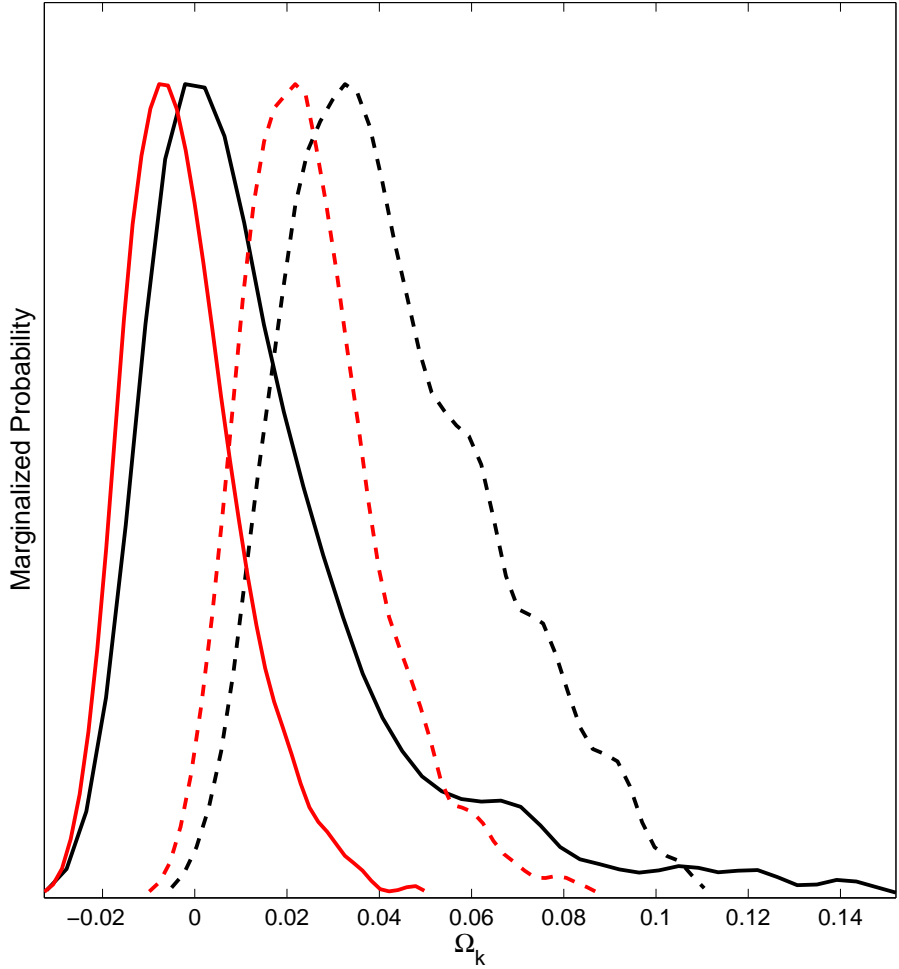}
\caption{The marginalized probabilities of $\Omega_k$. The solid lines denote the results
with $H_0=65$. The dashed lines denote the results with $H_0=72$. The black lines are for
the dark energy model $w_0+w_a z/(1+z)^2$ and the red lines are for the model
$w_0+w_a z/(1+z)$.}
\label{fig2}
\end{figure}

Since the angular scale of the sound horizon depends on the early
history of the Universe, so it strongly depends on $\Omega_r$.
However, we can neglect the effect of $\Omega_r$ when we evaluate
the distance modules $\mu(z)$ and the shift parameter $R$ because
the Universe is matter dominated. So only when we implement the CMB
data with $l_a$, we need to consider the effect of $\Omega_r$. We
know the energy density $\rho_r$ of radiation, so the dependence of
$\Omega_r=8\pi G\rho_r/(3 H_0^2)$ is manifested by the Hubble
constant $H_0$. Since we can neglect the effect of $\Omega_r$ in
fitting SN Ia data, so the effect of the observed value of $H_0$ can
be neglected by marginalizing over it. Therefore, we use the Hubble
constant $H_0$ as a free parameter instead of $\Omega_r$.
 The marginalized probabilities of $\Omega_k$ for
$H_0=65$ km/s/Mpc and $H_0=72$ km/s/Mpc are shown in Fig.
\ref{fig2}. We see that the results indeed depend on $H_0$. As
discussed in \citep{lar}, the combination of $R$ and $l_a$
approximates the WMAP3 data and the WMAP3 data depends on $H_0$
through $l_a$. So, as expected, $l_a$ also depends on $H_0$.  From
now on we also take $H_0$ as a fitting parameter, and impose a prior
of $H_0=72\pm 8$ km/s/Mpc \citep{freedman}. To understand why we can
marginalize over $H_0$ in fitting SN Ia data and treat $H_0$ as a
parameter in fitting WMAP3 data, we should think that we actually
treat $\Omega_r$, not $H_0$ as a parameter when fitting the WMAP3
data. The parameter $H_0$ is not the observed Hubble constant when
fitting the SN data because the normalization of the distance
modulus was chosen arbitrarily. In summary, we have six fitting
parameters for the DE models (\ref{lind}) and (\ref{wzeq}).

\section{Results}
In this section, we present our results. We first use the 182 gold
SN Ia data \citep{riess06}, then we use the ESSENCE data
\citep{riess06, essence,essence1}. For the SN Ia data, we consider
both the SN Ia flux averaging with marginalization over $H_0$
\citep{flux,flux1,flux2} and the analytical marginalization without
the flux averaging. The results with the analytical marginalization
are shown in solid lines and the results with flux averaging are
shown in dashed lines. We also put the $\Lambda$CDM model with the
symbol + in the contour plot.

\subsection{Gold SN Ia data}

\begin{figure}
\centering
\includegraphics[width=12cm]{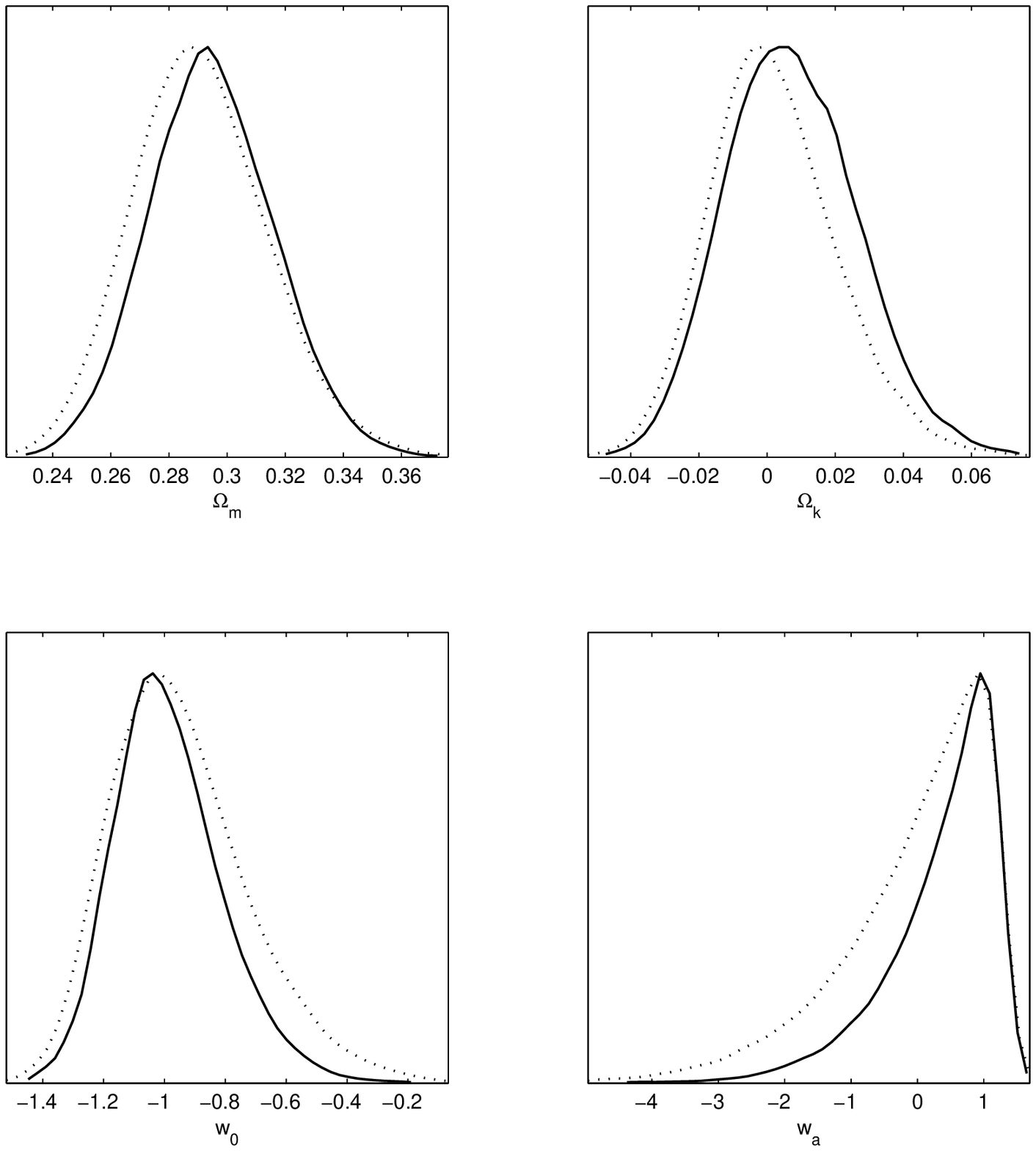}
\caption{The marginalized probabilities for the DE model $w_0+w_a z/(1+z)$
by using the gold SN Ia data. The solid lines denote the results
with analytical marginalization and the dashed lines denote the results with flux averaging.}
\label{fig3}
\end{figure}

\begin{figure}
\centering
\includegraphics[width=12cm]{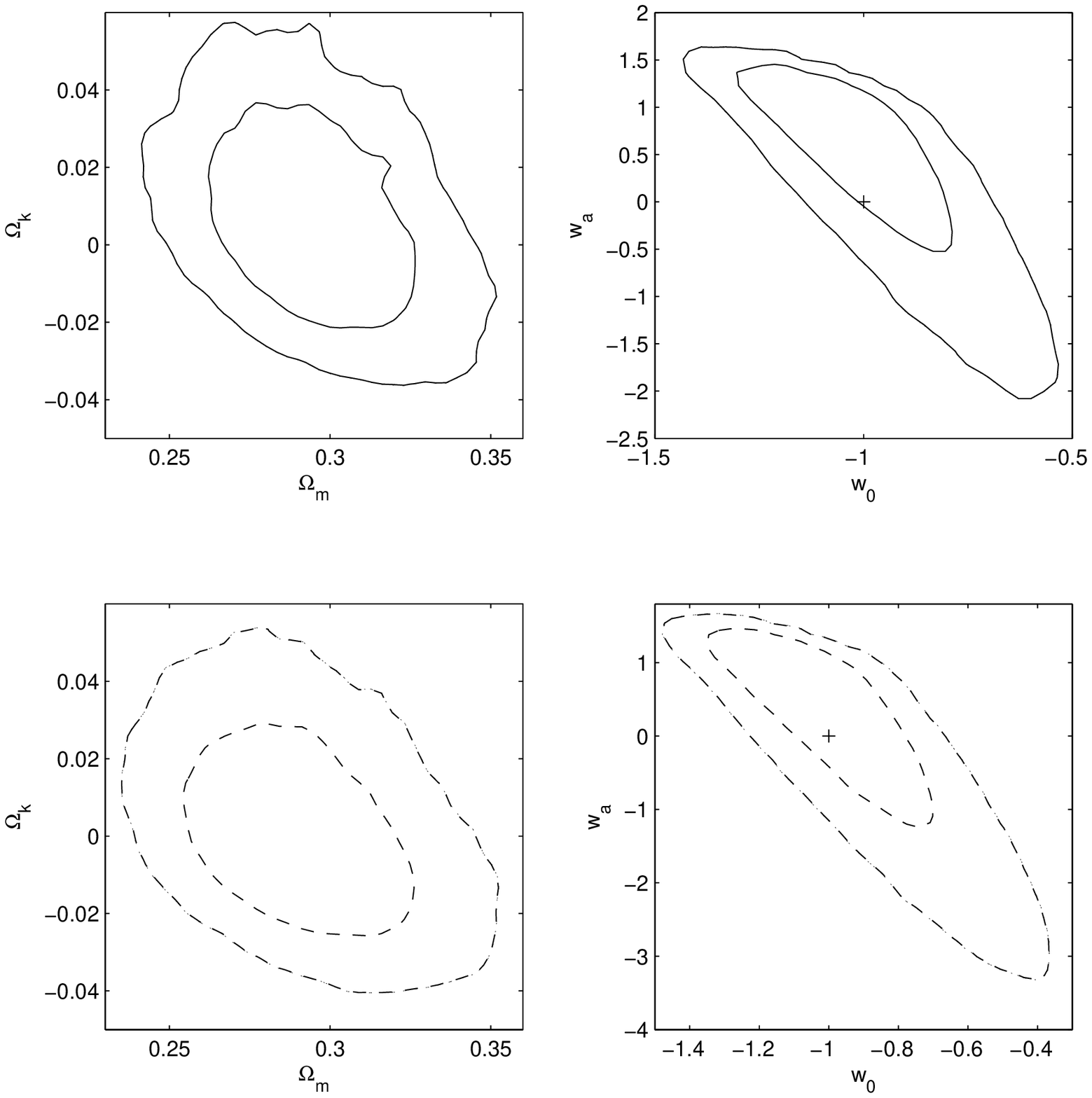}
\caption{The marginalized $1\sigma$ and $2\sigma$ $\Omega_m$-$\Omega_k$ and $w_0$-$w_a$ contours for the DE model $w_0+w_a z/(1+z)$
by using the gold SN Ia data.
The upper panels denote the results
with analytical marginalization and the lower panels denote the results with flux averaging.}
\label{fig4}
\end{figure}

Fig. \ref{fig3} shows the marginalized probabilities for $\Omega_m$,
$\Omega_k$, $w_0$ and $w_a$ for the DE model $w_0+w_a z/(1+z)$. Fig.
\ref{fig4} shows the marginalized $\Omega_m$-$\Omega_k$ and
$w_0$-$w_a$ contours. The $w_0$-$w_a$ contour with the flux
averaging is consistent with the result in \citet{wang07}. From
Figs. \ref{fig3} and \ref{fig4}, we see that the difference in the
results between the analytical marginalization and the flux
averaging is small. The $\Lambda$CDM model is consistent with the
observation at the $1\sigma$ level. The value of $w_a$ is better
constrained with the analytical marginalization.

\begin{figure}
\centering
\includegraphics[width=12cm]{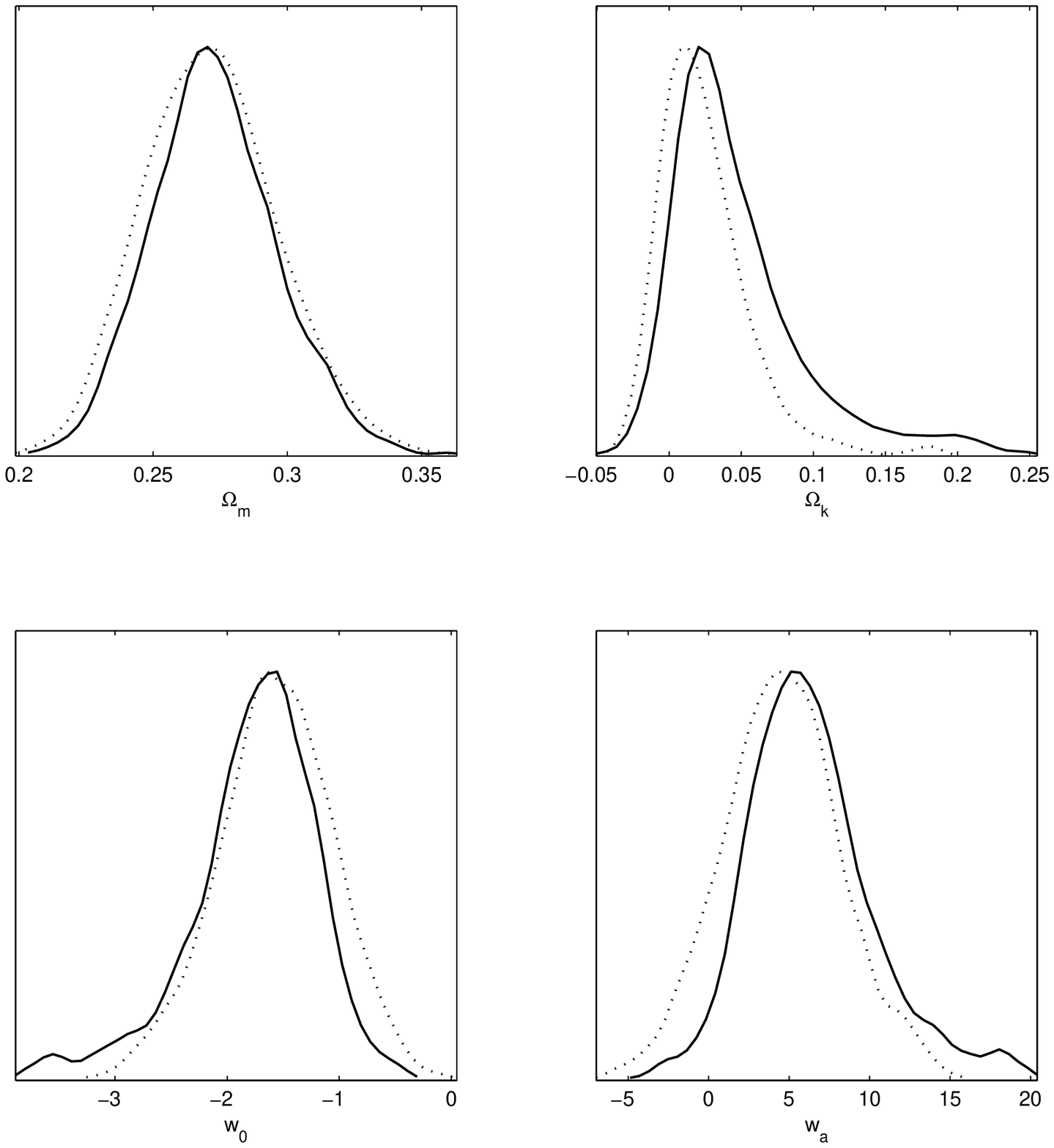}
\caption{The marginalized probabilities for the DE model $w_0+w_a z/(1+z)^2$
by using the gold SN Ia data. The solid lines denote the results
with analytical marginalization and the dashed lines denote the results with flux averaging.}
\label{fig5}
\end{figure}

\begin{figure}
\centering
\includegraphics[width=12cm]{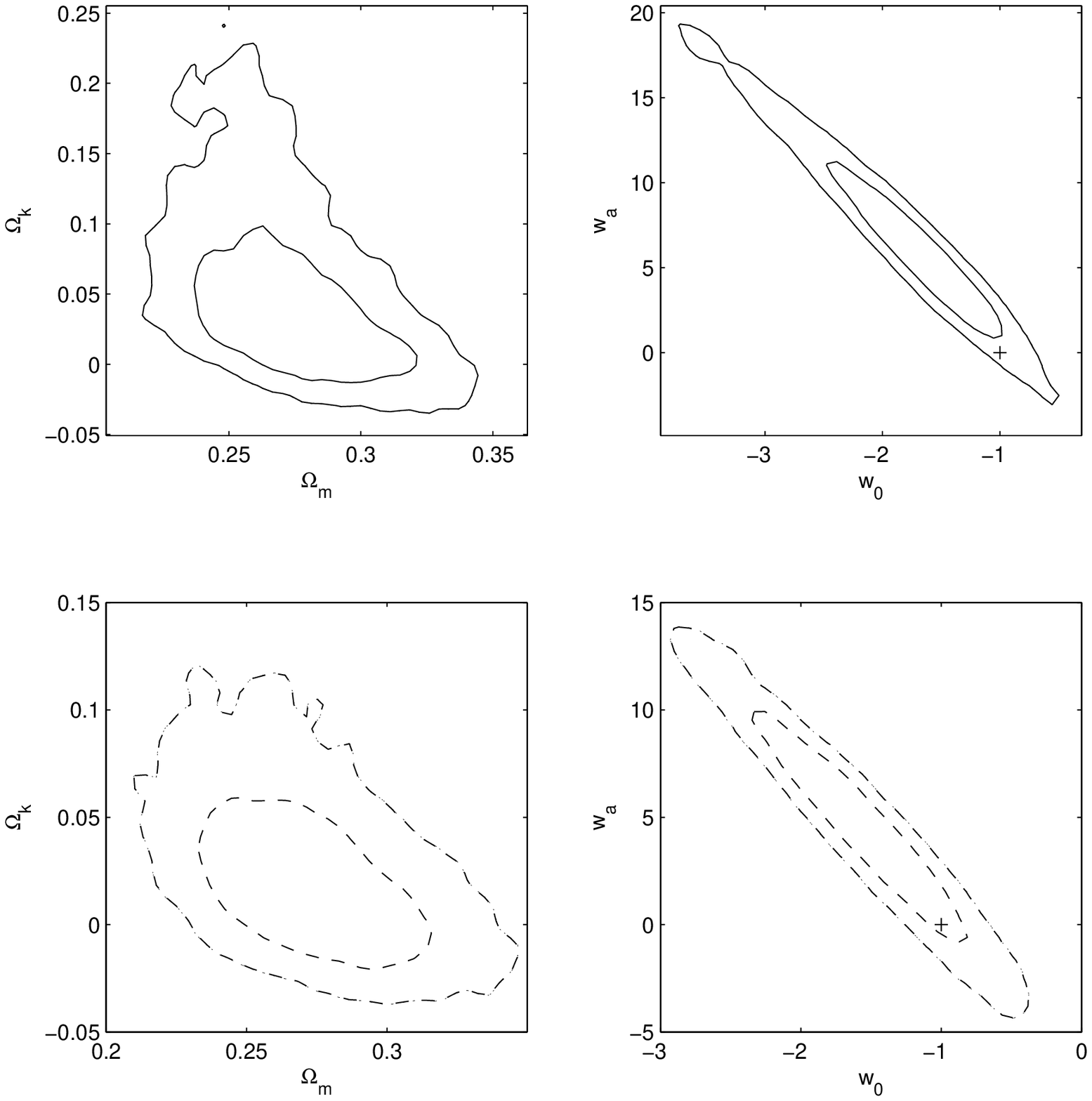}
\caption{The marginalized $1\sigma$ and $2\sigma$ $\Omega_m$-$\Omega_k$ and $w_0$-$w_a$ contours for the DE model $w_0+w_a z/(1+z)^2$
by using the gold SN Ia data.
The upper panels denote the results
with analytical marginalization and the lower panels denote the results with flux averaging.}
\label{fig6}
\end{figure}

Fig. \ref{fig5} shows the marginalized probabilities for $\Omega_m$,
$\Omega_k$, $w_0$ and $w_a$ for the DE model $w_0+w_a z/(1+z)^2$.
Fig. \ref{fig6} shows the marginalized $\Omega_m$-$\Omega_k$ and
$w_0$-$w_a$ contours.  From Figs. \ref{fig5} and \ref{fig6}, we see
that the parameters are a little better constrained with the flux
averaging. For the analytical marginalization, the $\Lambda$CDM
model is consistent with the observation at the $2\sigma$ level. For
the flux averaging, the $\Lambda$CDM model is consistent with the
observation at the $1\sigma$ level.

\subsection{ESSENCE data}

\begin{figure}
\centering
\includegraphics[width=12cm]{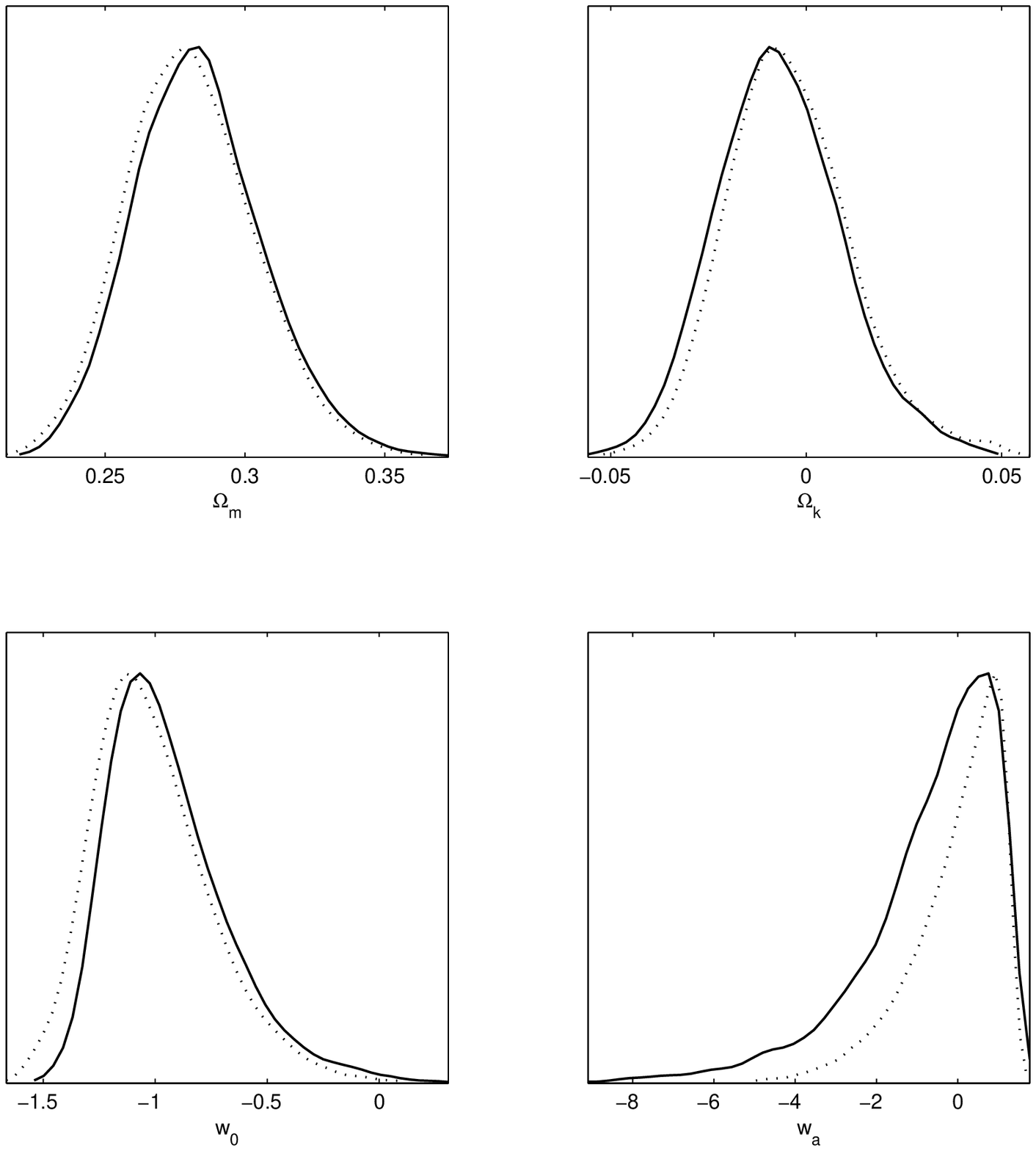}
\caption{The marginalized probability distributions for the dark energy model $w_0+w_a z/(1+z)$
by using the ESSENCE data.
The solid lines denote the results
without flux average and the dashed lines denote the results with flux average.}
\label{fig7}
\end{figure}

\begin{figure}
\centering
\includegraphics[width=12cm]{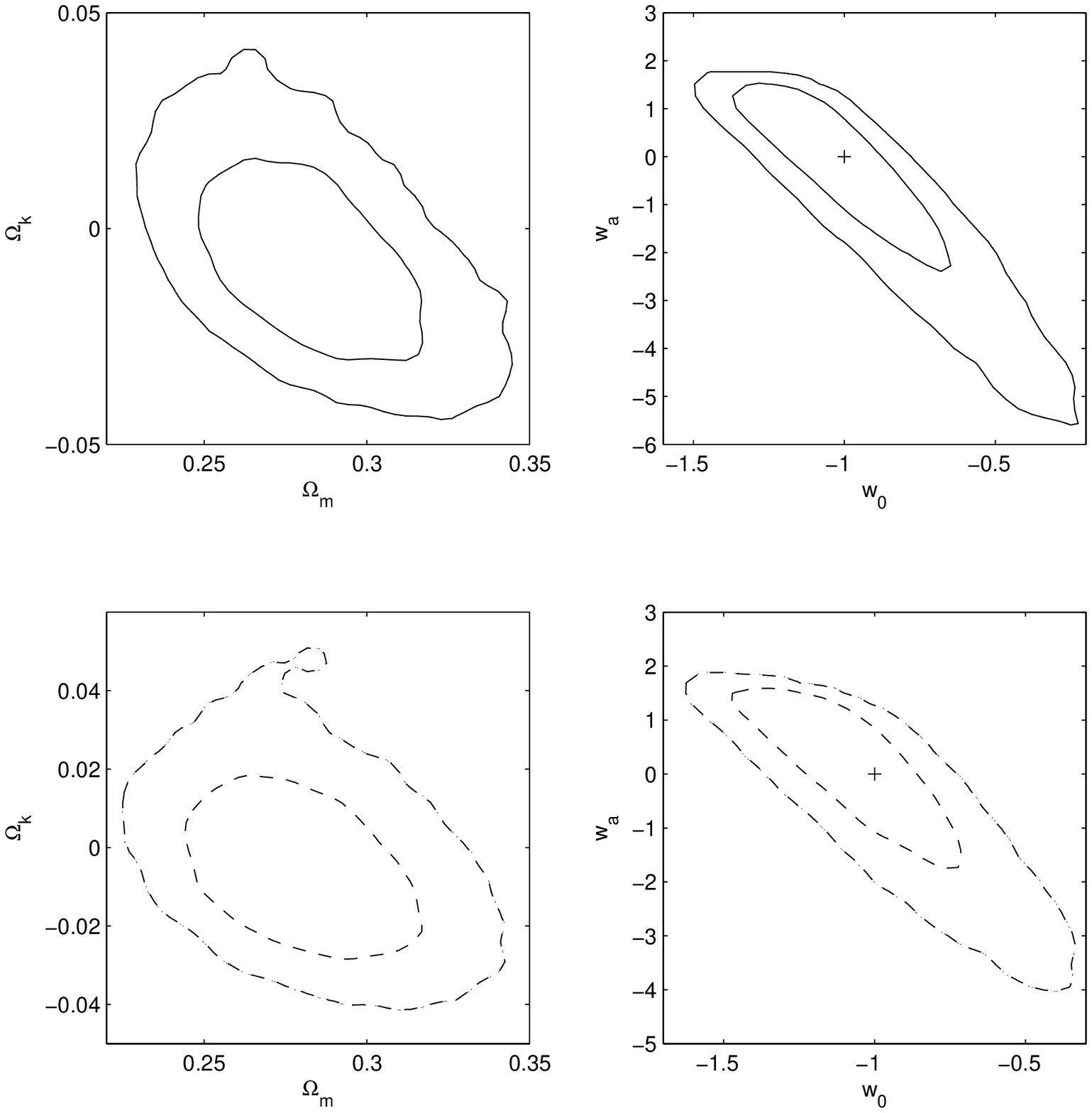}
\caption{The marginalized $\Omega_m$-$\Omega_k$ and $w_0$-$w_a$ contours for the dark energy model $w_0+w_a z/(1+z)$
by using the ESSENCE data.
The upper panels denote the results
without flux average and the lower panels denote the results with flux average.}
\label{fig8}
\end{figure}

Fig. \ref{fig7} shows the marginalized probabilities for $\Omega_m$,
$\Omega_k$, $w_0$ and $w_a$ for the DE model $w_0+w_a z/(1+z)$. Fig.
\ref{fig8} shows the marginalized $\Omega_m$-$\Omega_k$ and
$w_0$-$w_a$ contours. From Figs. \ref{fig7} and \ref{fig8}, we see
that the difference in the results between the analytical
marginalization and the flux averaging is small. The $\Lambda$CDM
model is consistent with the observation at the $1\sigma$ level.

\begin{figure}
\centering
\includegraphics[width=12cm]{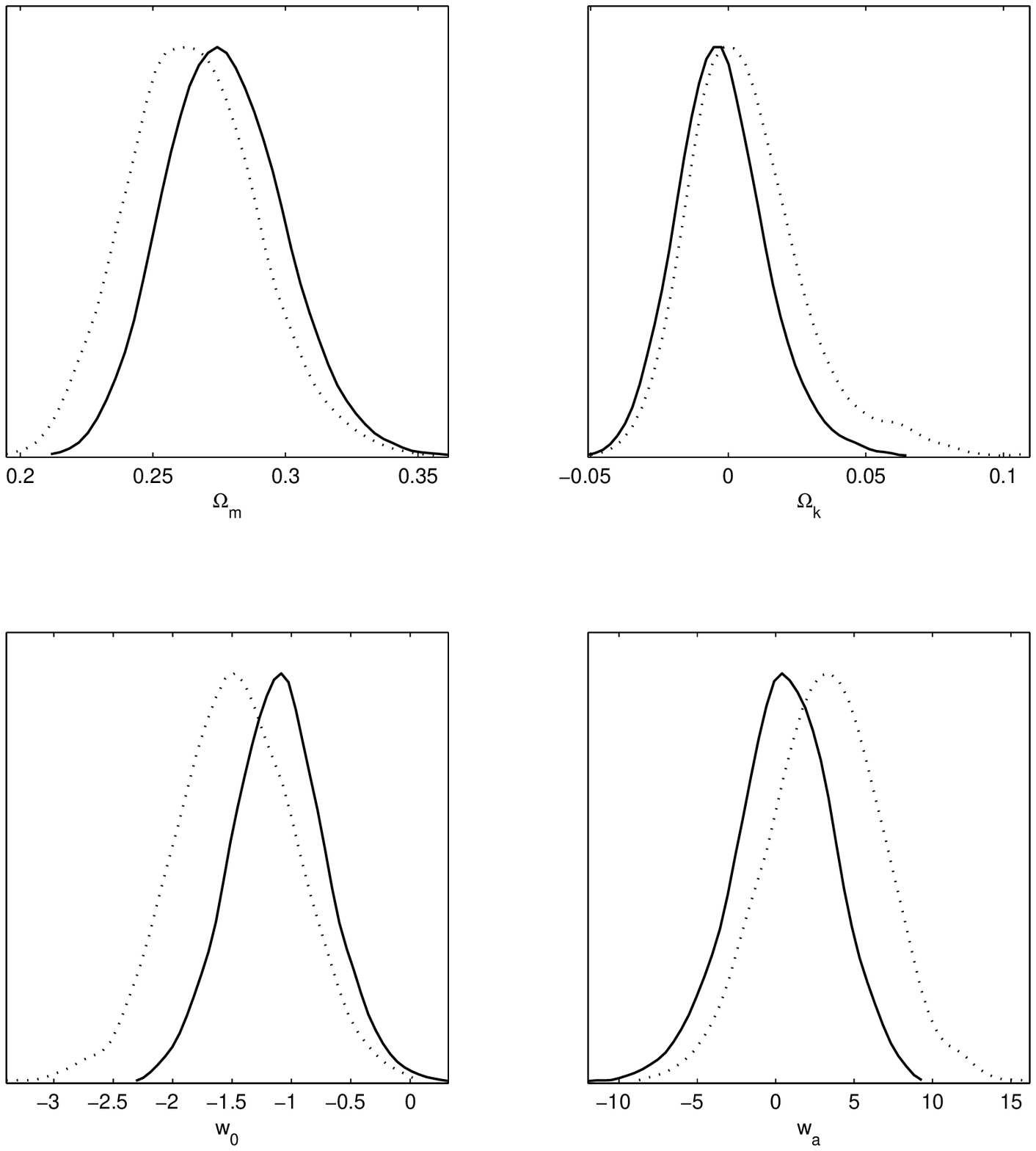}
\caption{The marginalized probability distributions for the dark energy model $w_0+w_a z/(1+z)^2$
by using the ESSENCE data.
The solid lines denote the results
without flux average and the dashed lines denote the results with flux average.}
\label{fig9}
\end{figure}

\begin{figure}
\centering
\includegraphics[width=12cm]{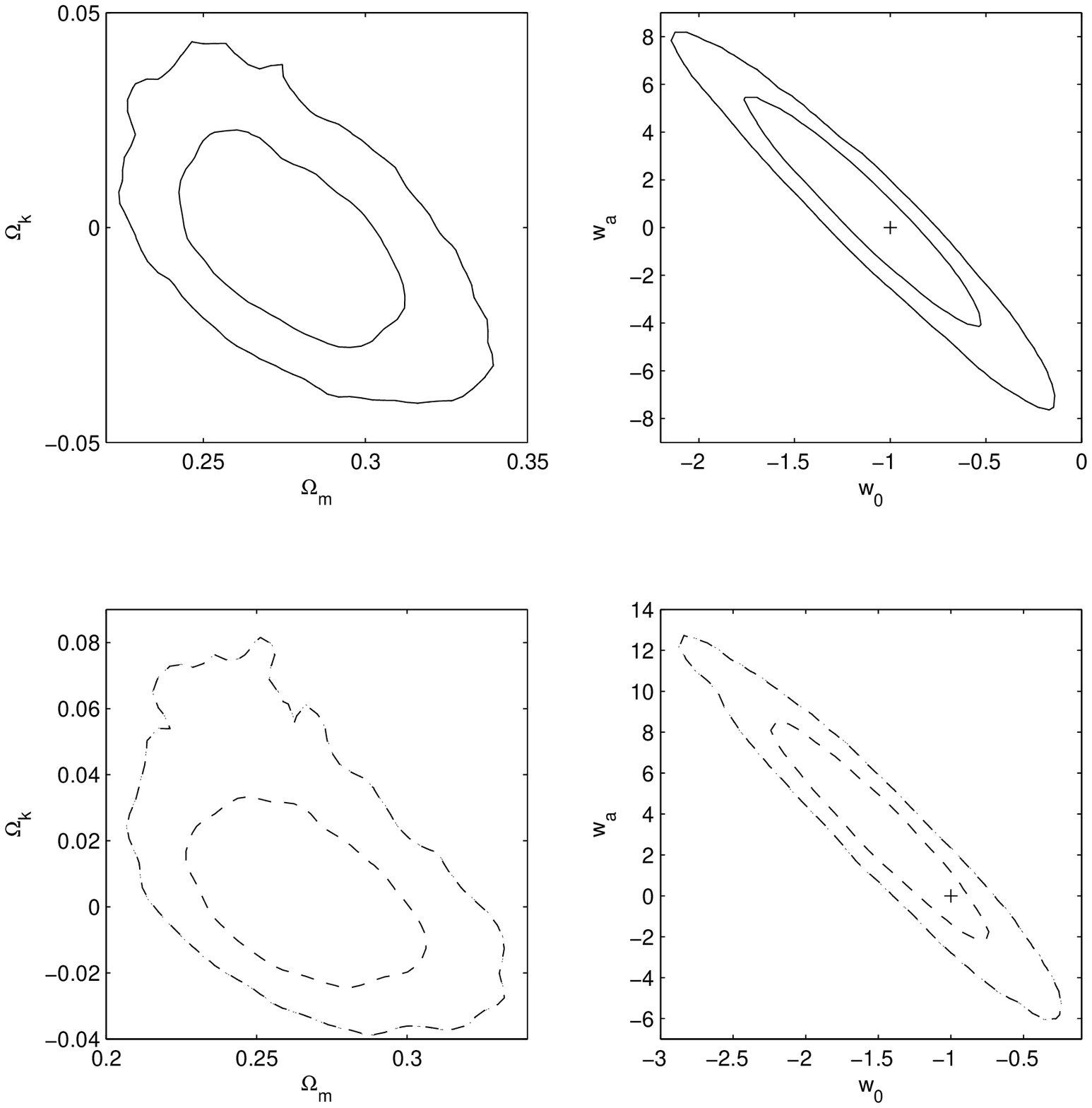}
\caption{The marginalized $\Omega_m$-$\Omega_k$ and $w_0$-$w_a$ contours for the dark energy model $w_0+w_a z/(1+z)^2$
by using the ESSENCE data.
The upper panels denote the results
without flux average and the lower panels denote the results with flux average.}
\label{fig10}
\end{figure}

Fig. \ref{fig9} shows the marginalized probabilities for $\Omega_m$,
$\Omega_k$, $w_0$ and $w_a$ for the DE model $w_0+w_a z/(1+z)^2$.
Fig. \ref{fig10} shows the marginalized $\Omega_m$-$\Omega_k$ and
$w_0$-$w_a$ contours.  From Figs. \ref{fig9} and \ref{fig10}, we see
that the parameters are a little better constrained with the
analytical marginalization. The $\Lambda$CDM model is consistent
with the observation at the $1\sigma$ level.

We summarize the results in Tables 1 and 2. We do not see much
improvement on the constraints on the DE parameters and the cosmic
curvature by using the flux averaging method. For the DE model
$w_0+w_a z/(1+z)$, the gold data gives better constraints than the
ESSENCE data on the DE parameters $w_0$ and $w_a$, but both data
give good constraints on the cosmic curvature. For the DE model
$w_0+w_a z/(1+z)^2$, the ESSENCE data gives much better constraint
on the cosmic curvature than the gold data, although the constraints
on the DE parameters $w_0$ and $w_a$ are almost the same for both
data. For the 182 gold data, the DE model $w_0+w_a z/(1+z)$ gives
much better constraints on the cosmic curvature $\Omega_k$. For the
ESSENCE data, the two DE models give almost the same constraint on
$\Omega_m$ and $\Omega_k$. For the DE model $w_0+w_a z/(1+z)$, the
mean value of $w_0$ determined from the observation tends to be
$w_0\geq -1$, while the mean value of $w_0$ is less than $-1$ for
the DE model $w_0+w_a z/(1+z)^2$.

From Tables 1 and 2, we see that the constraints on $\Omega_k$ are
almost the same for the two different DE models (\ref{lind}) and
(\ref{wzeq}). In other words, the results we obtained on $\Omega_k$
do not depend on the chosen models much. Recently, the authors in
\cite{clarkson} found that the assumption of a flat universe induces
critically large errors in reconstructing the dark energy equation
of state at $z\ga 0.9$ even if the true cosmic curvature is very
small, $\Omega_k\sim 0.01$ or less. They obtained the result by
fitting the data derived from a DE model with $\Omega_k\neq 0$ with
a flat model, so the result may not be conclusive. To see how the
value of $\Omega_k$ affect the constraints on the property of DE, we
perform the MCMC analysis on the DE models (\ref{lind}) and
(\ref{wzeq}) with $\Omega_k=0$. The results are reported in Tables 3
and 4. Although the uncertainties of $\Omega_k$ change the values of
$w_0$ and $w_a$, the ranges of $w_0$ and $w_a$ are almost the same
for small $\Omega_k$. 

In conclusion, we first confirm previous results that the shift
parameter $R$ alone does not give good constraint on $\Omega_k$, we
must combine $R$ and $l_a$ to constrain $\Omega_k$. By using $R$,
$l_a$ and their covariance matrix, we get almost the same results as
those obtained by using the original WMAP3 data. Without calculating
the power spectrum, the fitting process is much faster and
efficient. The cosmic curvature is found to be $|\Omega_k| \la
0.03$.

\acknowledgments
 YGG and AW thank Yun Wang for the help with the
MCMC method. YGG is grateful of Zong-hong Zhu for fruitful
discussions, and he is supported by NNSFC under grant No. 10605042.
A. Wang is partially supported by the VPR funds, Baylor University.

\appendix
\section{Analytical marginalization on $H_0$}

By assuming a flat prior $P(H_0)=1$ for $H_0$, the marginalization
over $H_0$ means
\begin{equation}
\label{intmarg}
L=e^{-\chi^2_m/2}=\int e^{-\chi^2/2}P(H_0)dH_0=\int e^{-\chi^2/2}dH_0.
\end{equation}
Let $x=5\log_{10}H_0$ and
$\alpha_i=\mu_{obs}(z_i)-25-5\log_{10}[(1+z_i){\rm
sinn}(\sqrt{|\Omega_k|}\int_0^{z_i} dz'/E(z'))/\sqrt{|\Omega_k|}]$,
and substitute Eq. (\ref{chi}) into the above Eq. (\ref{intmarg}),
we get
\begin{eqnarray}
\label{intmarg1}
\begin{array}{cl}
L&=\frac{\ln 10}{5}\int dx\exp\left[-\frac{1}{2}\sum_i\frac{(\alpha_i+x)^2}{\sigma^2_i}+\frac{\ln 10}{5}x\right]\\
&=\frac{\ln 10}{5}\int
dx\exp\left[-\frac{1}{2}\left(\sum_i\frac{1}{\sigma_i^2}\right)
\left(x+\frac{\sum_i\alpha_i/\sigma_i^2-\ln
10/5}{\sum_i1/\sigma_i^2}\right)^2\right.\\
&\left.\ \ \ -\frac{1}{2}\sum_i\frac{\alpha^2_i}{\sigma^2_i}
+\frac{1}{2}\frac{(\sum_i\alpha_i/\sigma_i^2-\ln 10/5)^2}{\sum_i1/\sigma_i^2}\right]\\
&=\frac{\ln
10}{5}\left(\frac{2\pi}{\sum_i1/\sigma_i^2}\right)^{1/2}\exp(-\frac{1}{2}\sum_i\frac{\alpha^2_i}{\sigma^2_i}
+\frac{1}{2}\frac{(\sum_i\alpha_i/\sigma_i^2-\ln
10/5)^2}{\sum_i1/\sigma_i^2}).
\end{array}
\end{eqnarray}
So the minimum $\chi^2$ is
\begin{equation}
\label{minchi2}
\chi^2_m=\sum_i\frac{\alpha^2_i}{\sigma^2_i}-\frac{(\sum_i\alpha_i/\sigma_i^2-\ln
10/5)^2}{\sum_i1/\sigma_i^2} -2\ln\left(\frac{\ln
10}{5}\sqrt{\frac{2\pi}{\sum_i1/\sigma_i^2}}\right).
\end{equation}

\clearpage

\begin{table}
\label{table1}
 \caption{The marginalized results with $1\sigma$
errors for the model $w_0+w_a z/(1+z)$}
\begin{tabular}{|ccccc|}
\hline
&  \multicolumn{2}{c} {Gold Data}  &\multicolumn{2}{c|} {Essence Data} \\
& Analytical & Flux & Analytical & Flux\\ \hline
$\Omega_m$&$0.29^{+0.03}_{-0.02}$ & $0.29\pm 0.02$ & $0.28^{+0.03}_{-0.02}$ & $0.28\pm 0.02$ \\  \hline
$\Omega_k$& $0.007^{+0.023}_{-0.019}$ & $0.002\pm 0.018$ & $-0.007\pm 0.016$ & $-0.004^{+0.015}_{-0.016}$\\ \hline
$w_0$& $-0.99^{+0.18}_{-0.16}$ & $-0.95\pm 0.22$ & $-0.94\pm 0.25$ & $-1.0^{+0.24}_{-0.26}$ \\   \hline
$w_a$& $0.34\pm 0.77$ & $-0.05^{+1.04}_{-1.09}$& $-0.70^{+1.54}_{-1.52}$& $-0.26^{+1.29}_{-1.31}$  \\ \hline
\end{tabular}
\end{table}

\begin{table}
\label{table2} \caption{The marginalized results with $1\sigma$
errors for the model $w_0+w_a z/(1+z)^2$}
\begin{tabular}{|ccccc|}
\hline
&  \multicolumn{2}{c} {Gold Data}  &\multicolumn{2}{c|} {Essence Data} \\
& Analytical & Flux & Analytical & Flux \\ \hline
$\Omega_m$& $0.27^{+0.03}_{-0.02}$ & $0.27^{+0.03}_{-0.02}$& $0.28^{+0.02}_{-0.03}$ & $0.29\pm 0.02$ \\  \hline
$\Omega_k$& $0.05\pm 0.04$& $0.02^{+0.03}_{-0.02}$& $-0.002^{+0.015}_{-0.016}$& $-0.013\pm 0.011$ \\ \hline
$w_0$& $-1.8^{+0.6}_{-0.5}$& $-1.6^{+0.6}_{-0.5}$& $-1.1\pm 0.4$& $-1.1^{+0.4}_{-0.5}$ \\   \hline
$w_a$& $6.4\pm 3.6$ & $4.5\pm 3.6$& $0.5\pm 3.1$& $0.6\pm 3.3$ \\ \hline
\end{tabular}
\end{table}

\begin{table}
\label{table3}
 \caption{The marginalized results with $1\sigma$
errors for the model $w_0+w_a z/(1+z)$ with $\Omega_k=0$}
\begin{tabular}{|ccccc|}
\hline
&  \multicolumn{2}{c} {Gold Data}  &\multicolumn{2}{c|} {Essence Data} \\
& Analytical & Flux & Analytical & Flux\\ \hline
$\Omega_m$&$0.29 \pm 0.02$ & $0.29 \pm 0.02$ & $0.27^{+0.02}_{-0.01}$ & $0.27\pm 0.02$ \\  \hline
$w_0$& $-0.98^{+0.17}_{-0.13}$ & $-0.98\pm 0.20$ & $-1.06^{+0.18}_{-0.17}$ & $-1.09 \pm 0.22$ \\   \hline
$w_a$& $0.28^{+0.65}_{-0.64}$ & $0.15 \pm 0.78$& $0.17^{+0.79}_{-0.85}$& $0.28 \pm 0.87$  \\ \hline
\end{tabular}
\end{table}

\begin{table}
\label{table4} \caption{The marginalized results with $1\sigma$
errors for the model $w_0+w_a z/(1+z)^2$ with $\Omega_k=0$}
\begin{tabular}{|ccccc|}
\hline
&  \multicolumn{2}{c} {Gold Data}  &\multicolumn{2}{c|} {Essence Data} \\
& Analytical & Flux & Analytical & Flux \\ \hline
$\Omega_m$& $0.29 \pm 0.02$ & $0.28\pm 0.02$& $0.27 \pm 0.02$ & $0.27\pm 0.02$ \\  \hline
$w_0$& $-1.23\pm 0.26$& $-1.22^{+0.31}_{-0.30}$& $-1.15^{+0.29}_{-0.30}$& $-1.29^{+0.33}_{-0.32}$ \\   \hline
$w_a$& $2.28^{+1.75}_{-1.71}$ & $1.94^{+1.98}_{-1.96}$& $1.01^{+2.15}_{-2.14}$& $1.87^{+2.11}_{-2.12}$ \\ \hline
\end{tabular}
\end{table}

\end{document}